\documentclass[aps,prl,twocolumn,groupedaddress,showpacs]{revtex4}
\usepackage{graphicx}

\begin{document}

\title{Prediction of genomic properties and classification of life by protein length
distributions}

\author{Dirson Jian Li}
\email[]{dirson@mail.xjtu.edu.cn}
\author{Shengli Zhang}

\affiliation{Department of Applied Physics, Xi'an Jiaotong
University, Xi'an 710049, China}

\begin{abstract}
Much evolutionary information is stored in the fluctuations of
protein length distributions. The genome size and non-coding DNA
content can be calculated based only on the protein length
distributions. So there is intrinsic relationship between the coding
DNA size and non-coding DNA size. According to the correlations and
quasi-periodicity of protein length distributions, we can classify
life into three domains. Strong evidences are found to support the
order in the structures of protein length distributions.

\end{abstract}


\maketitle
\date{}

\section{Introduction}

There is an analogy between the current status of particle physics
and molecular biology, in each of which there is a theoretical
framework to explain the particle world or living world (the
Standard Model for the former while the Central Dogma and gene
regulation for the latter), but we do not know the dynamical
mechanisms that underlie these theoretical frameworks. It is
generally believed that the genetic information is stored in DNA
sequences. This often results in a misapprehension that all the
information about a species is stored in the sequences of its
genome. Actually, only a part of the information of life is stored
in the sequences, from which we may infer the structures and
functions of macromolecules. But there is still other information
concerns the underlying mechanism of the evolution of life, which
can not be acquired by analyzing the sequences. For example, the
mechanism that determines the protein length distribution differs
from the mechanism that determines in the protein sequences, which
requires new explanation other than the current theoretical
framework.

The protein length distribution is an intrinsic properties of a
species that concerns the underlying mechanism of molecular
evolution. But the nature of the protein length distribution is
still unknown. Someone considered it as a result of stochastic
process, while others noticed the order in the length distributions
\cite{Protein length distributions1}\cite{Protein length
distributions2}\cite{Protein length distributions3}\cite{Protein
length distributions4}. We found that much information of the
evolution of life is stored in the fluctuations of protein length
distributions. The genome size of a species, even its coding DNA
size and non-coding DNA size can be calculated by the protein length
distribution of this species. We also found that the three domains
of life (Bacteria, Archaea, and Eucarya) can be classified based on
the correlations or quasi-periodicity of protein length
distributions; furthermore, we can obtain the phylogeny of the three
domains \cite{3domains}. These results shows that the protein length
distributions are not random, the fluctuations in the distribution
are intrinsic properties and can be taken as the fingerprint of a
certain species. We propose a linguistic model to explain the
generation of protein sequences and consequently try to explain the
nature of protein length distributions. The model indicates that the
complexity of the grammars that generate protein sequences is
related to the biological complexity of the species. The protein
length distributions might be taken as clues to discover the
underlying mechanism of molecular evolution.

\section{Prediction of genome size and non-coding DNA content
by protein length distributions.}

The evolution of genome size is one of the central problems in the
study of molecular evolution \cite{Gregory}\cite{Gregory2}. We can
write down the experimental formulae of genome size and the ratio of
non-coding DNA to coding DNA based on the protein length
distributions, which may help us understand the mechanism of genome
size evolution.

The protein length distribution of a species $\alpha$ can be denoted
by a vector
\begin{equation}{\mathbf x(\alpha)}=(x_1(\alpha), x_2(\alpha), ..., x_k(\alpha),
...),\end{equation} where there are $x_k(\alpha)$ proteins in its
entire proteome whose lengths are $k$ amino acids (a.a.). Our data
of the protein length distributions are obtained from the data of
$n=106$ complete proteomes ($n_b=85$ bacteria, $n_a=12$ archaea,
$n_e=7$ eukaryotes and $n_v=2$ viruses) in the database Predictions
for Entire Proteomes (PEP) \cite{PEP}. The total protein length
distribution of all the species in PEP is $\mathbf X=\sum_{\alpha
\in PEP} \mathbf x(\alpha)$, from which we found that there are few
proteins longer than $m=3000$ a.a. in the complete proteomes. We can
neglect them and set $m$ as the cutoff of protein length in our
calculations. Hence the vectors $\mathbf x(\alpha)$ can be reduced
to $m$-dimensional vectors to represent the protein length
distributions.

\begin{figure}
\centering{
\includegraphics[width=60mm]{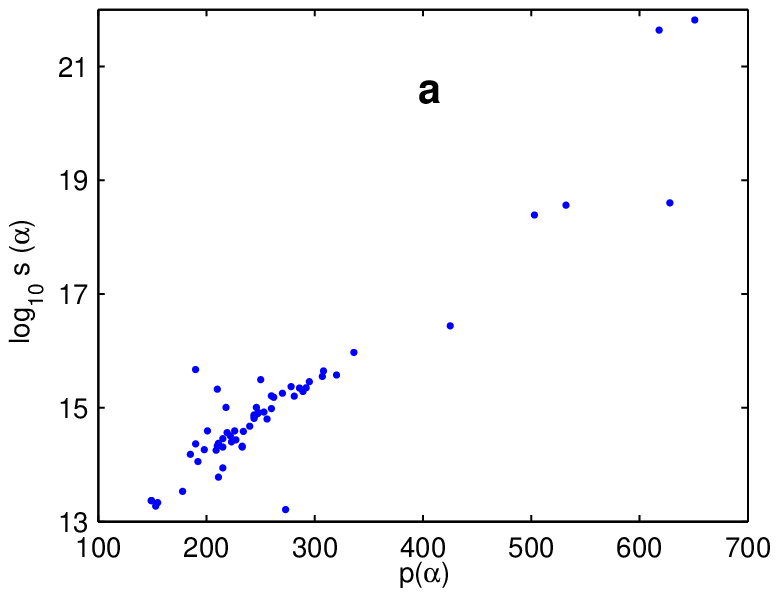}
\includegraphics[width=60mm]{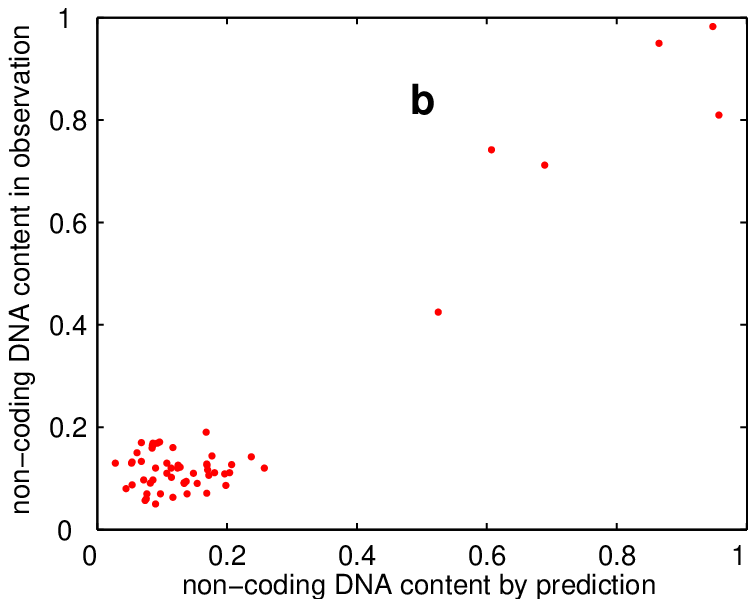}}
\label{fig1} \caption{\small {\bf Prediction of genome size and
non-coding DNA content by the protein length distribtuions.} {\bf
a,} The relationship between logarithm of genome size $\log
s(\alpha)$ and the number of peaks $p(\alpha)$ in the protein length
distributions. {\bf b,} The predictions of non-coding DNA contents
by Eqn. 2 agree with the non-coding DNA contents of species.}
\end{figure}

A peak at protein length $l$ in the fluctuations of protein length
distribution of species $\alpha$ can be denoted by $x_l(\alpha)$,
which is required to be greater than both $x_{l-1}(\alpha)$ and
$x_{l+1}(\alpha)$. The number of peaks in the fluctuations of
protein length distribution of species $\alpha$ can be denoted by
$p(\alpha)$. We found that there is an exponential relationship
between genome size $s(\alpha)$ and number of peaks $p(\alpha)$ (Fig
1a)
\begin{equation}s(\alpha)=s'\exp(\frac{p(\alpha)}{p_0}),\end{equation}
where $s'=8.36\times10^4$ base pairs (bp) and $p_0=70.6$.

We had also obtained another formula to calculate the genome size
\cite{Dirson_Cambrian}
\begin{equation}s(\alpha)=7.96 \times
10^6\exp(\frac{\eta(\alpha)}{0.165}-\frac{\theta(\alpha)}{0.176}),
\end{equation} where $\eta(\alpha)$ is the non-coding DNA content
in the complete genome of species $\alpha$ and the correlation polar
angle is defined by
$\theta(\alpha)=\frac{2}{\pi}\arccos(\frac{\mathbf
x(\alpha)}{\left|\mathbf x(\alpha)\right|} \cdot \frac{\mathbf
X}{\left|\mathbf X\right|})$. In terms of the above two experimental
formulae of genome zise, we obtained a formula to calculate the
non-coding DNA content:
\begin{equation}\eta(\alpha)=0.752+0.938\ \theta(\alpha)-0.00234\
p(\alpha),\end{equation} where both $\theta$ and $p$ depend only on
protein length distributions. There are $59$ species in PEP whose
non-coding DNA contents are known according to to Ref. \cite{eta}.
The prediction of the non-coding DNA contents agrees with the
experimental observations (Fig 1b).

Thus, we found that the genomic properties can be predicted based
only on the protein length distributions. The information of genome
size $s(\alpha)$, non-coding DNA size $\eta(\alpha) s(\alpha)$ and
coding DNA size $(1-\eta(\alpha))s(\alpha)$ are stored in the
fluctuations of protein length distribution of species $\alpha$.
More peaks in the fluctuations indicates lager genome size. And the
number of peaks in the fluctuations also relates to the non-coding
DNA content. The protein length distribution concerns only the
coding DNA, but we can deduce the non-coding DNA size by such a
distribution. This shows that there must be a universal mechanism to
adjust the ratio of coding DNA and non-coding DNA in each species.

In some studies, the protein length are considered as random
variable of a stochastic process. If so, there should not be close
relationship between the number of peaks $p(\alpha)$ and the genome
size $s(\alpha)$; and the outline of the protein length distribution
would be more smooth when the genome size increases. Our results
show that the protein length is not a random variable of stochastic
process and the fluctuations are intrinsic properties.

The number of peaks of the fluctuations in the protein length
distribution $p(\alpha)$ is an intrinsic property of a species
$\alpha$. We also found that the correlation between $p(\alpha)$ and
$s(\alpha)$ is closer than the correlation between $p(\alpha)$ and
$\bar{l}(\alpha)$ (Fig 1b and Fig 2c). It is suggested that the
biological complexity is related to the non-coding DNA content, and
the biological complexity is also related to the genome size for
prokaryotes \cite{21}\cite{22}\cite{23}. The number of peaks
$p(\alpha)$ can be interpreted as the complexity of structures of
protein length distribution. The linguistics plays significant roles
in the organization of protein or DNA sequences
\cite{ProteinLinguistice}\cite{Searls}. We proposed a linguistic
mechanism to account for the generation of protein sequences
\cite{Dirson_HolographicPrinciple}. The bell-shaped outline and the
fluctuations of the protein length distribution can be simulated by
a linguistic model. The number of peaks of the fluctuations in the
protein length distribution is determined by the complexity of
grammars. The correlation between $p(\alpha)$ and $s(\alpha)$
indicates a relationship between the complexity of the structure of
protein length distribution and the biological complexity of that
species, both of which may result from the complexity of grammars in
the sequences.

\section{Classification of life by correlation and quasi-periodicity
of protein length distributions.}

\begin{figure}
\centering{
\includegraphics[width=60mm]{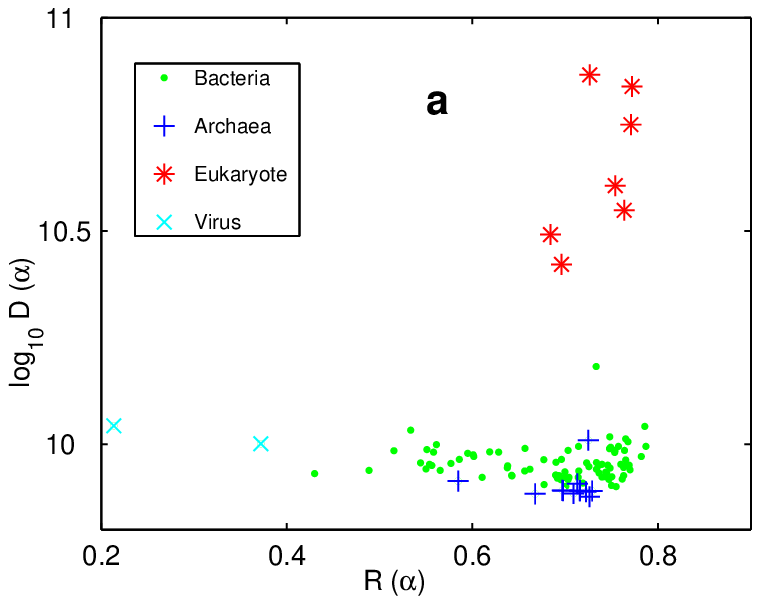}
\includegraphics[width=60mm]{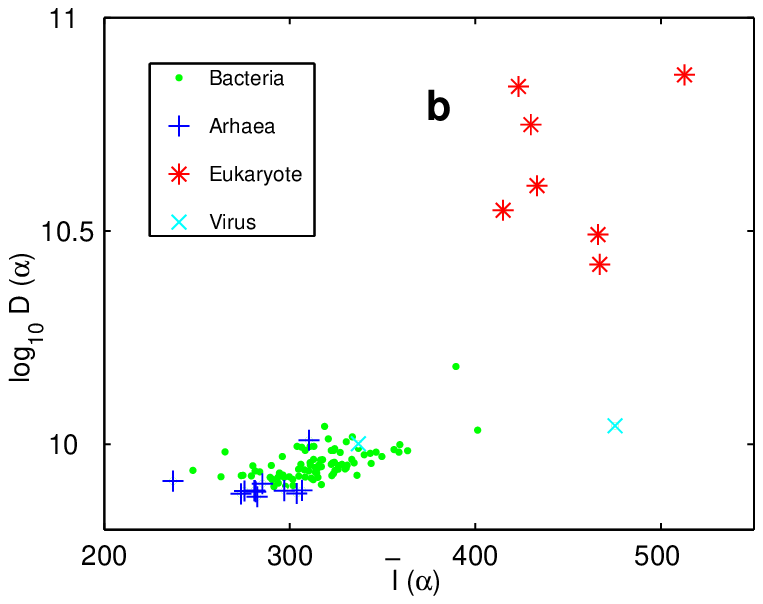}
\includegraphics[width=60mm]{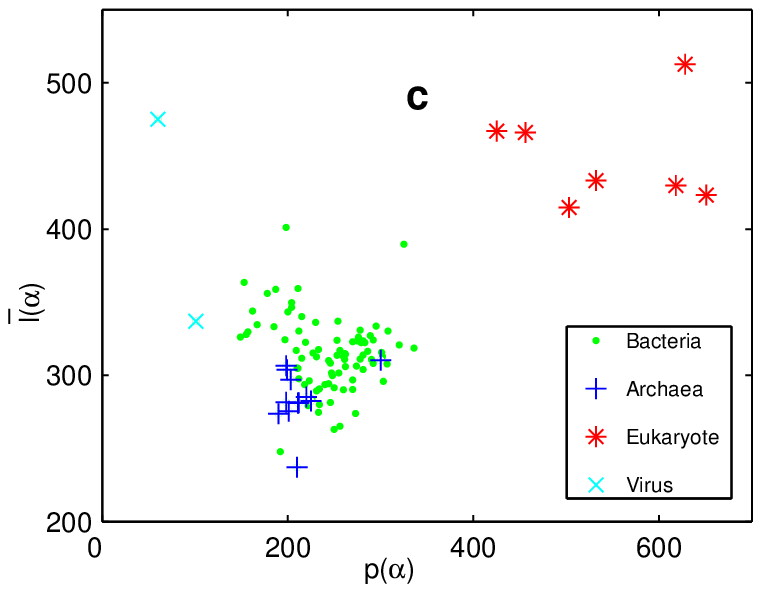}
\includegraphics[width=60mm]{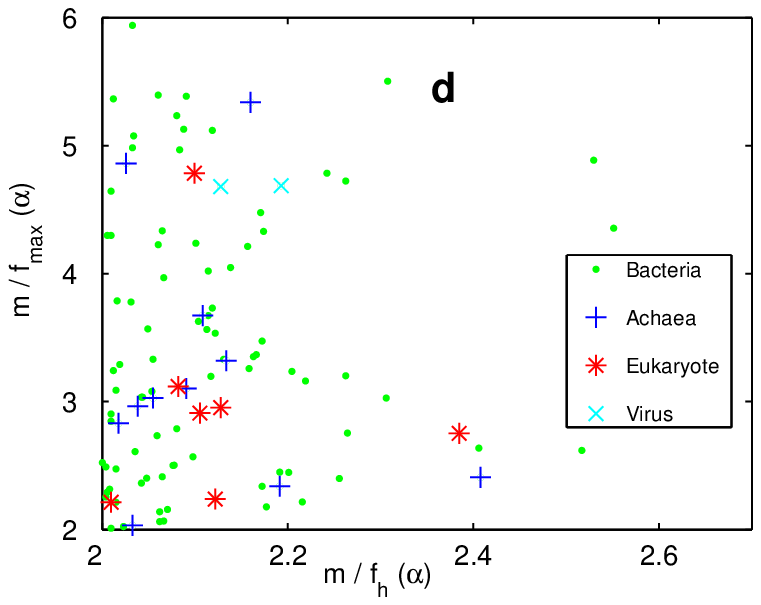}}
\label{fig1} \caption{\small {\bf Clustering analysis of life based
on the correlation and quasi-periodicity of protein length
distributions.} The species in three domains cluster together in
different areas in $R-\log D$, $\bar{l}-\log D$ and $\bar{l}-p$
planes respectively. {\bf a,} Clustering analysis by average
correlation $R(\alpha)$ and average Minkowski distance $D(\alpha)$.
{\bf b,} Clustering analysis by average protein length
$\bar{l}(\alpha)$ and average Minkowski distance $D(\alpha)$. {\bf
c,} The relationship between average protein length
$\bar{l}(\alpha)$ and the number of peaks $p(\alpha)$. {\bf d,} The
relationship between long period $m/f_h(\alpha)$ and short period
$m/f_{max}(\alpha)$ in the protein length distributions. We can
observe parallel structures in the distributions of species.}
\end{figure}

Molecular sequence analysis provides a more precise and profound
method in classification of life than classical taxonomy. Based upon
rRNA sequence comparisons, life on this planet can be divided into
three domains: the Bacteria, the Archaea, and the Eucarya
\cite{3domains}. The differences that separate the three domains are
of a more profound nature than the differences that separate
classical five kingdoms (Monera, Protista, Fungi, Plantae,
Animalia). The correlations and quasi-periodicity in the protein
length distributions may originate in a general underlying mechanism
of generation of protein sequences \cite{periodicity1}
\cite{periodicity2}. According to the theory of multivariation data
analysis, we can classify life in the three domains based only on
the protein length distributions.

The correlation coefficient of protein length distributions between
species $\alpha$ and $\beta$ is defined by \begin{equation}
r(\alpha,\beta)=\frac{\sum_{k=1}^m
(x_k(\alpha)-\bar{x}({\alpha}))(x_k(\beta)-\bar{x}({\beta}))}{\sqrt{\sum_{k=1}^m
(x_k(\alpha)-\bar{x}({\alpha}))^2}\sqrt{\sum_{k=1}^m
(x_k(\beta)-\bar{x}({\beta}))^2}},\end{equation} where
$\bar{x}({\alpha})=\frac{1}{m}\sum_{k=1}^m x_k(\alpha)$. And the
average correlation coefficient of species $\alpha$ can be defined
by \begin{equation}R(\alpha)=\frac{1}{106}\sum_{\beta \in PEP}
r(\alpha,\beta).\end{equation} We can also define the Minkowski
distance between species $\alpha$ and $\beta$ as
\begin{equation}d(\alpha,\beta)=(\sum_{k=1}^m
\left|\frac{x_k(\alpha)}{\left|\left|\mathbf
x(\alpha)\right|\right|}-\frac{x_k(\beta)}{\left|\left|\mathbf
x(\beta)\right|\right|}\right|^q)^{1/q},\end{equation} where we
chose the parameter $q=1/4$ in the calculation. And the average
Minkowski distance of species $\alpha$ can be defined by
\begin{equation}D(\alpha)=\frac{1}{106}\sum_{\beta \in PEP}
d(\alpha,\beta).\end{equation} The average correlation coefficient
$R(\alpha)$ and the average Minkowski distance $D(\alpha)$ represent
the evolutionary relationship between species $\alpha$ and all the
other species. The more the average correlation coefficient is (or
inversely the less the average Minkowski distance is), the closer
the evolutionary relationship is.

We can classify life by clustering analysis based only on the
protein length distributions. In the $R-\log D$ plane, we found that
the species in the three domains cluster together respectively (Fig
2a). The similar results can also be obtained by other methods. The
average protein length in the proteome of species $\alpha$ can be
calculated by protein length distribution:
\begin{equation}\bar{l}(\alpha)=\frac{\sum_{k=1}^m k\
x_k(\alpha)}{\sum_{k=1}^m x_k(\alpha)}.\end{equation} According to
the distributions of species in the plots of $\bar{l}-R$,
$\bar{l}-\log D$ and $\bar{l}-p$, we found that the species in the
three domains also cluster together in the corresponding plots
respectively.

\begin{figure}
\centering{
\includegraphics[width=80mm]{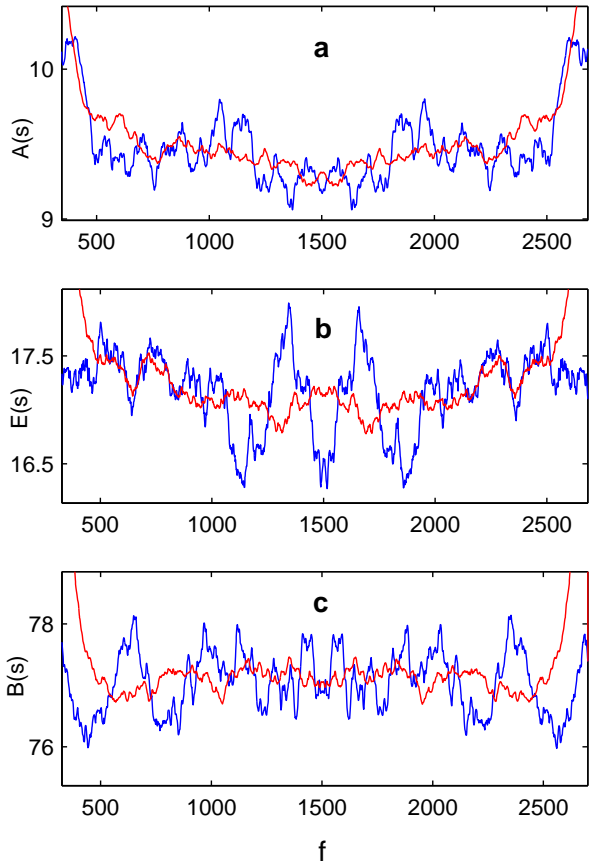}}
\label{fig1} \caption{\small {\bf Differences in the average Fourier
spectra of the protein length distributions for three domains.} The
ranges for averaging are chosen as $s1=100$ (Blue lines) and
$s2=300$ (Red lines) for the three domains respectively. {\bf a,}
The total average Fourier spectra $A(s1)$ and $A(s2)$ for Archaea.
{\bf b,} The total average Fourier spectra $E(s1)$ and $E(s2)$ for
Eucarya. {\bf c,} The total average Fourier spectra $B(s1)$ and
$B(s2)$ for Bacteria.}
\end{figure}

We can also classify life according to the quasi-periodicity of
protein length distributions. There are underlying orders in the
organization of protein sequences, so we can observe the
quasi-periodic structures in the protein length distributions.
According to Fourier analysis, we can also observe the clustering of
species for different domains.

The abstract discrete fourier transformation of the protein length
distribution $\mathbf x(\alpha)$ is:
\begin{equation} \mathbf{y}_f(\alpha) = \frac{1}{\sqrt{m}} \left|\left| \sum _{k=1}
^m x_k(\alpha) e^{2 \pi i (k-1)(f-1)/m} \right|\right|.
\end{equation} The frequency of the highest peak in the
fluctuations of the spectrum $\mathbf{y}(\alpha)$ can be denoted by
$f_h(\alpha)$. The maximum frequency for the top $10$ highest peaks
in the fluctuations of the spectrum $\mathbf{y}(\alpha)$ can be
denoted by $f_{max}(\alpha)$.

We found that there is an interesting relationship between the
highest peak frequency $f_h(\alpha)$ and the average protein length
$\bar{l}$ of species (Fig. 3 in Ref.
\cite{Dirson_HolographicPrinciple}). The distribution of species in
$f_h-\bar{l}$ plane shows a regular pattern: the species in the
three domains gathered in three rainbow-like arches respectively.
This pattern strongly indicates the intrinsic correlation among the
protein length distributions, which can not achieve if the protein
length distributions are stochastic. The periodicity of the protein
length distributions can be inferred by the correlations between
$f_h$ and $f_{max}$. The regular distribution of species in Fig. 2d
indicates the correlation between the short period $m/f_{h}$ and the
long period $m/f_{max}$. Therefore, the quasi-periodic structures of
the spectra indicate the correlations between long proteins and
short proteins in a proteome.

The total spectra for three domains Bacteria, Archaea and Eucarya
are $ \mathbf b=\frac{1}{n_b}\sum_{\alpha \in Bacteria} \mathbf
y(\alpha)$, $\mathbf a=\frac{1}{n_a}\sum_{\alpha \in Archaea}
\mathbf y(\alpha)$ and $\mathbf e=\frac{1}{n_e}\sum_{\alpha \in
Eucarya} \mathbf y(\alpha)$ respectively. So the average spectra for
the three domains are defined as follows respectively:
\begin{eqnarray}
\mathbf B (s)=\frac{1}{2s+1}\sum_{k=f-s}^{f+s}b_k\\
\mathbf A (s)=\frac{1}{2s+1}\sum_{k=f-s}^{f+s}a_k\\
\mathbf E (s)=\frac{1}{2s+1}\sum_{k=f-s}^{f+s}e_k,\end{eqnarray}
where $f=1+s, ..., m-s$ and we chose $s1=100$ and $s2=300$ as the
ranges for averaging in calculations. We found that the outlines of
$\mathbf A$ and $\mathbf E$ are similar, both of which have convex
bottoms (Fig 3a, b), but they are dissimilar with the outline of
$\mathbf B$, which has a concave bottom (Fig 3c). According to the
phylogeny of the three domains, the relationship between Archaea and
Eucarya is more closer than the relationship between Archaea and
Bacteria. So the outlines of the average spectra of the three
domains reveal the phylogeny of the three domains.

\section{Conclusion and discussion}

We can conclude that much evolutionary information is stored in the
protein length distributions, according to which we can calculate
the genome size and non-coding DNA content. The mechanism that
determines the protein length distribution concerns not only coding
DNA but also non-coding DNA. So we demonstrated the intrinsic
relationship between coding DNA size and non-coding DNA size. New
methods are proposed to classify life into three domains based only
on the protein length distributions. We confirm that there are
quasi-periodic structures in the protein length distributions, and
we found the relationship between average protein length and the
frequencies of Fourier transformation of protein length
distributions.

The protein length distributions can not be considered as random
fluctuations. We found strong evidences of the underlying mechanism
in the organization of amino acids in protein sequences, which
indicates the languages in the protein sequences.  We proposed a
linguistic model to accord for the protein length distribution,
which suggested that there is a close relationship between the
complexity of grammars of the protein sequences and the biological
complexity of the species.

\end{document}